\documentclass[12pt]{iopart}
\usepackage{graphicx}
\usepackage{bm} 
\usepackage{iopams} 
\usepackage[pdftex]{hyperref} 

 \newcommand{\iotabar}{\mbox{$\,\iota\!\!$-}}
 \newcommand{\Eqn}[1]{\eref{eq:#1}} 
 
 \newcommand{\be}{\begin{eqnarray}}
 \newcommand{\ee}{\end{eqnarray}}

 \newcommand{\pr}{p_{\rm r}}
 \newcommand{\qr}{q_{\rm r}}
 \renewcommand{\d}{d}

\begin{document}

\title[Generalised action-angle coordinates]{Generalised action-angle coordinates defined on island chains}

\author{R L Dewar, S R Hudson and A M Gibson}

\address{Research School of Physics and Engineering, The Australian University, Canberra 0200, Australia}
\address{Princeton Plasma Physics Laboratory, Princeton NJ, USA}
\ead{robert.dewar@anu.edu.au}
\begin{abstract}
Straight-field-line coordinates are very useful for representing magnetic fields in toroidally confined plasmas, but fundamental problems arise regarding their definition in 3-D geometries because of the formation of islands and chaotic field regions, ie non-integrability. 
In Hamiltonian dynamical systems terminology these coordinates are similar to action-angle variables, but these are normally defined only for integrable systems.
In order to describe 3-D magnetic field systems, a generalisation of this concept was proposed recently by the present authors that unified the concepts of ghost surfaces and quadratic-flux-minimising (QFMin) surfaces. This was based on a simple canonical transformation generated by a change of variable $\theta = \theta(\Theta,\zeta)$, where $\theta$ and $\zeta$ are poloidal and toroidal angles, respectively, with $\Theta$ a new poloidal angle chosen to give pseudo-orbits that are a) straight when plotted in the $\zeta,\Theta$ plane and b) QFMin pseudo-orbits in the transformed coordinate. These two requirements ensure that the pseudo-orbits are also c) ghost pseudo-orbits. In the present paper, it is demonstrated that these requirements do not \emph{uniquely} specify  the transformation owing to a relabelling symmetry. A variational method of solution that removes this lack of uniqueness is proposed.
\end{abstract}

\submitto{\PPCF}
\maketitle


\section{\label{sec:intro}Introduction}
Recent calculations \cite{Hudson_Breslau_08} of heat diffusion along chaotic field lines show that the isotherms correspond very closely with the ``approximate'' magnetic surfaces, associated with magnetic island chains, known as ghost surfaces \cite{Hudson_Dewar_96}. These surfaces include the ``X-point'' and ``O-point'' closed field lines of their associated islands. (By ``O-point''  field line we mean either the elliptically stable field line at the center of an island or its hyperbolically unstable continuation if it has undergone a period-doubling bifurcation.)
Closed field lines make the magnetic action stationary, the hyperbolic X-point field lines in the chaotic separatrices being minima and the O-point field lines being minimax or saddle points of the action. Ghost surfaces are constructed by interpolating smoothly between these two closed-field-line classes by evolving the O-point field lines into the X-point field lines along paths of steepest descent of action, thus generating a family of ``pseudo-orbits,'' i.e. paths that come close to making the action stationary.

Ghost surfaces have nice mathematical properties but are difficult to construct and have no obvious physical interpretation. An alternative approach to defining approximate magnetic surfaces passing through magnetic islands, is to use the quadratic-flux-minimizing (QFMin) surfaces introduced by Dewar, Hudson and Price \cite{Hudson_Dewar_96,Dewar_Hudson_Price_94}. These surfaces have the computational attraction of being easy to construct using (pseudo) field-line tracing methods, and the physical attraction of being defined in terms of a measure of the magnetic flux transport through the surface, but have been found to exhibit undesirable distortions in some circumstances. Thus a unified approach that combines the best features of ghost and QFMin surfaces is desirable.

Both QFMin surfaces and ghost surfaces can be formulated in terms of the action gradient, but the action gradient is \emph{coordinate dependent}. We have recently \cite{Dewar_Hudson_Gibson_10} exploited this coordinate dependence by finding the conditions under which a transformation from a given ``old'' poloidal coordinate $\theta$ to a ``new'' poloidal angle $\Theta$ makes ghost and QFMin surfaces identical (a process we call \emph{reconciliation}). An added benefit of this construction is that it makes the pseudo-orbits straight when plotted in the $\zeta,\Theta$ Cartesian plane. As this is similar to the way the action-angle transformation for integrable systems makes the true orbits rectilinear we term this a \emph{generalised action-angle transformation}.

Our principal motivation is finding an optimal generalisation of straight-field-line magnetic coordinates in toroidal systems. As it is well known (see e.g. \cite{Hudson_Dewar_09} and references therein) that magnetic fields can be described as $1\frac{1}{2}$-degree-of-freedom Hamiltonian systems we shall build our treatment upon standard classical mechanics, as for instance in \cite{Goldstein_80}, with the poloidal angle $\theta$ as the generalised coordinate and the toroidal angle $\zeta$ as the ``time''. Also, in this paper we use only the Lagrangian approach to classical mechanics as it is somewhat simpler in the single-torus case we study. The translation between the Lagrangian and Hamiltonian approaches in the context of our generalised pseudo-orbit approach is developed in \cite{Dewar_Hudson_Gibson_12}.

An analogue of our reconciliation prescription has recently been implemented \cite{Gibson_12} for a discrete-time dynamical system, an iterated area-preserving map (the standard or Chirikov--Taylor map), as a model problem. A variational approach was used to perform numerical experiments aimed at finding whether the prescription can reconcile ghost and QFMin almost-invariant curves at arbitrary nonlinearity. (\emph{Unreconciled} ghost and QFMin curves have been constructed for standard-map nonlinearity parameters up to $k = 100$ \cite{Dewar_Khorev_95}.) Reconciliation transformations were successfully constructed for quite high nonlinearity ($k \sim 1$) after it was realised that there was a lack of uniqueness in the prescription of \cite{Dewar_Hudson_Gibson_10} that could be fixed by reducing the number of Fourier basis functions appropriately.

In the present paper we identify this lack of uniqueness as due to a \emph{relabelling symmetry}. Relabelling symmetries also occur in MHD and fluid dynamics \cite{Webb_Zank_07} and are analogous to gauge symmetries in physical field theories \cite{Guida_ZinnJustin_08}. Thus, making the new poloidal angle unique is analogous to fixing a gauge to make the representation of a field unique. We propose a dual-objective-functional variational method for constructing a unique reconciliation transformation.

In \sref{sec:CT} we review the basic classical mechanics concepts required and introduce the concept of periodic pseudo-orbits and associated almost-invariant tori. In \sref{sec:QFMinGhost} we give a concise derivation of the reconciliation conditions found in \cite{Dewar_Hudson_Gibson_10} and in \sref{sec:relabtrnsfn} we demonstrate that, given one solution satisfying the reconciliation conditions, an infinity of solutions may be generated by relabelling the points at which the pseudo-orbits cross the $\zeta = 0$ surface of section. A primary objective functional that respects the relabelling symmetry and, when minimised to zero, gives valid reconciliation transformations is presented in \sref{sec:VarQFMinGhost}. Approaches for fixing the non-uniqueness problem are discussed in \sref{sec:2ndOpt} where we propose a secondary objective function, whose minimisation with respect to reconciliation transformations generated by relabelling transformations will both remove (or at least reduce) the non-uniqueness and ensure invertibility of the reconciliation transformation. Some areas for further research are briefly indicated in \sref{sec:Conclusion}.

\section{A simple canonical transformation}\label{sec:CT}

A \emph{$(\pr,\qr)$-periodic path} is defined in the $\theta,\zeta$ plane by the curve $\theta = \vartheta(\zeta)$ subject to the periodicity condition $\vartheta(\zeta + 2\pi \qr) = \vartheta(\zeta) + 2\pi \pr$ ($\pr$ and $\qr$ mutually prime integers). Physical examples of such paths are the elliptic and hyperbolic closed field lines (periodic orbits) passing through the O and X points of a magnetic island chain formed through the resonant destruction of a rational surface with rotational transform $\iotabar = \pr/\qr$ (safety factor $q = \qr/\pr$), but we also consider pseudo-orbits --- paths that are ``not quite'' physical. 

As our theory is based on variational principles, we also consider \emph{variations} $\delta\vartheta$ of paths away from either physical or pseudo-orbits. For instance, the \emph{action integral} $S$ defined on an arbitrary $(\pr,\qr)$-periodic path is defined as a functional of the path function $\vartheta$ by the integral
\begin{equation}
	S[\vartheta]=\int^{2\pi q}_0 L(\vartheta,\vartheta'(\zeta),\zeta)\, d \zeta \;,
	\label{eq:actiondef}
\end{equation}
where $L \equiv L(\theta,\dot\theta,\zeta)$ is the \emph{Lagrangian}.
Varying $\vartheta$ in \eref{eq:actiondef} and integrating by parts we find the \emph{functional derivative} of $S$ as the coefficient of $\delta\vartheta$ in $\delta S$,
\begin{equation}
	\frac{\delta S}{\delta\theta} = L_{\theta} - \frac{d L_{\dot\theta}}{d\zeta} \;,
	\label{eq:actiongrad}
\end{equation}
where $L_{\theta}$ and $L_{\dot\theta}$ denote the partial derivatives of $L$ with respect to its first and second arguments, respectively. In the following we refer to $\delta S/\delta\theta$ as the \emph{action gradient} as it can be thought of as the generalisation of the gradient of a function to the infinite-dimensional space of path functions $\vartheta(\zeta)$. The action gradient $\delta S/\delta\theta$ can also be shown to play the role of a phase-space flux density (or magnetic flux density in the case of field-line dynamics.)

Hamilton's Principle \cite{Goldstein_80} is the statement that $S$ is stationary ($\delta S/\delta\theta = 0$) on physical orbits, i.e. the \emph{true equation of motion is obtained by setting the action gradient to zero}. We shall term paths for which the action gradient is not zero, but in some sense small, \emph{pseudo-orbits}.

We also term toroidal surfaces composed of families of pseudo-orbits \emph{almost-invariant tori}, specific cases being  \emph{ghost tori} when the pseudo-orbit families are constructed by an action-gradient flow joining true orbits, and \emph{QFMin tori} when the pseudo-orbit families are constructed variationally to minimise the ``quadratic flux'' $\frac{1}{2}\int\!\!\!\!\!\!\!\int\,(\delta S/\delta\theta)^2 d\theta d\zeta$, which is a measure of flux transport through the almost-invariant tori. The Euler--Lagrange equation for QFMin pseudo-orbits is $(d/d\zeta)\delta S/\delta\theta = 0$, implying that the action gradient is constant along each QFMin pseudo-orbit (which includes the case of a true orbit, when the constant is zero). For details see \cite{Dewar_Hudson_Gibson_12}.

The momentum canonically conjugate to $\theta$ is
\begin{equation}
	\label{eq:Idef}
	I = L_{\dot\theta} \;.
\end{equation}
Following \cite{Dewar_Hudson_Gibson_10} we shall seek to reconcile the QFMin and ghost formulations on a \emph{single} almost-invariant torus by using a canonical transformation generated by a change of generalised coordinate from $\theta$ to a new poloidal angle $\Theta$,
\be \theta = \theta(\Theta,\zeta) \;,
	\label{eq:thetaTransfn}
\ee
transforming the Lagrangian by the  point transformation \cite[pp 33, 386]{Goldstein_80} $L_{\rm new}(\Theta,\dot\Theta,\zeta) = L_{\rm old}(\theta,\dot\theta,\zeta)$. (As this is simply a statement that the Lagrangian is invariant under the transformation,  henceforth we leave the subscripts, ``old'' and ``new'' implicit as it is clear from the arguments which is meant.) We assume that the mapping $\Theta \mapsto \theta$ is invertible, implying that $\theta(\Theta,\zeta)$ is a monotonic (increasing) function of $\Theta$ for all $\zeta$.

Equation  \eref{eq:thetaTransfn} generates a \emph{canonical transformation} $\theta,I \mapsto \Theta,J$, where the new momentum conjugate to $\Theta$ is $J \equiv \partial L/\partial\dot\Theta$. Differentiating \eref{eq:thetaTransfn} we find
\be
\dot\theta = (\partial_{\zeta} + \dot\Theta\,\partial_{\Theta})\theta(\Theta,\zeta)
	\equiv \theta_{\zeta} + \dot{\Theta}\,\theta_{\Theta} \;,
	\label{eq:thetadot}
\ee
where $\theta_{\Theta}$ and $\theta_{\zeta}$ denote the partial derivatives of the transformation function $\theta$ with respect to its first and second arguments, respectively. (Invertibility implies that $\theta_{\Theta} > 0$.) Thus $L = L(\theta,\theta_{\zeta} + \dot{\Theta}\,\theta_{\Theta},\zeta$), giving
\be
	J = \theta_{\Theta}L_{\dot\theta}
	\equiv \theta_{\Theta}I \;.
	\label{eq:Jdef}
\ee
Note that this canonical transformation deforms the \emph{whole} phase space by \emph{rescaling} the momentum variable by an  amount that varies with $\theta$. While this is rather drastic, it does not matter for our current purposes as we are only interested in the neighbourhood of a single phase-space torus.

As the action along any path is invariant under the transformation, the variation $\delta S$ is also invariant, implying $\delta\Theta\,\delta S/\delta\Theta = \delta\theta\,\delta S/\delta\theta$ for all $\delta\theta$. Noting that, from \eref{eq:thetaTransfn}, $\delta\theta = \theta_{\Theta}(\Theta,\zeta) \delta\Theta$, we immediately see that the action gradients in the new and old variables are related by
\be \frac{\delta S}{\delta\Theta} = \theta_{\Theta}\frac{\delta S}{\delta\theta} \;.
	\label{eq:newActionGrad}
\ee

\section{Reconciliation of QFMin and Ghost surfaces}\label{sec:QFMinGhost}

The Euler--Lagrange equation \cite{Dewar_Hudson_Gibson_12} deriving from the quadratic flux minimisation (QFMin) principle in the new coordinate [varying the family of paths in $\Theta,\zeta$ space making up a trial torus, with a given transformation function $\theta(\zeta,\Theta)$] is
\be \frac{d}{d\zeta} \frac{\delta S}{\delta\Theta} = 0
	\;. \label{eq:QFMinEL}
\ee
That is,  the action gradient $\delta S/\delta\Theta$ is constant on each individual member of the family of pseudo-orbits that makes up the $(\pr,\qr)$ almost-invariant  torus under consideration. Denoting this constant by $\nu$ we have
\be \frac{\delta S}{\delta\Theta} = \nu(\Theta_0)
	\;, \label{eq:QFMinELint}
\ee
where $\Theta_0$ is the initial value of $\Theta$ on a pseudo-orbit, which we here use as a pseudo-orbit label. The function $\nu(\Theta_0)$ is constant along each pseudo-orbit but varies in an oscillatory fashion across the pseudo-orbit family, passing through zero at the action-minimizing and minimax true orbits \cite{Dewar_Khorev_95}.

Ghost pseudo-orbits are defined in the new coordinates by a gradient flow driven by the action gradient,
\be	\frac{D\Theta}{D T}  =  \frac{\delta S}{\delta\Theta} 
	\;, \label{eq:ThetaLaggradflow}
\ee
where the evolution variable $T$ is a label for ghost pseudo-orbits which goes from $-\infty$ to $+\infty$ or \emph{vice versa} depending on whether the evolution is up or down the action gradient. (We need to consider both cases to fill in the gaps between action-minimising and minimax orbits, and we use the notation $DT$ to emphasise that this flow is \emph{across} the family of pseudo-orbits rather than along them like the pseudo-dynamics generated by $d/d\zeta$.)

By reconciliation we mean that these two classes of pseudo-orbits are equivalent, so $T$ and $\Theta_0$ are functionally dependent: $T = T(\Theta_0)$, $DT = T'(\Theta_0)D\Theta_0$.
Eliminating $\delta S/\delta\Theta$ between \Eqn{QFMinELint} and \Eqn{ThetaLaggradflow} we find the  \emph{reconciliation condition}
\be \frac{D\Theta}{D\Theta_0}  = T'(\Theta_0)\nu(\Theta_0) \;,
	\label{eq:reconcilcon}
\ee
which puts an important constraint on the \emph{reconciliation transformation} \eref{eq:thetaTransfn}: \emph{it must be such that $D\Theta/D\Theta_0$ is independent of $\zeta$}. (Assuming this can be satisfied, it then relates the ghost pseudo-orbit evolution parameter $T$ to the QFMin pseudo-orbit label $\Theta_0$.) But, at $\zeta = 0$, $D\Theta/D\Theta_0 \equiv 1$ by definition. Thus the reconciliation condition implies $D\Theta/D\Theta_0 \equiv 1$ for all $\zeta$.

This condition implies that $\Theta(\zeta|\Theta_0)$ must separate in the form $\Theta_0 + f(\zeta)$, with $f$ arbitrary except for periodicity requirements. The simplest and most natural choice to try is to take $f$ linear in $\zeta$,
\be
	\Theta(\zeta|\Theta_0) = \Theta_0 + \iotabar\zeta \;,
	\label{eq:SFL}
\ee
with $\iotabar \equiv p/q$. Equation \eref{eq:SFL} conjugates the $\theta$ pseudo-dynamics to rigid rotation,
\be
	\vartheta(\zeta|\theta_0) = \theta(\Theta_0 + \iotabar\zeta,\zeta) \;, \:
	\theta_0 \equiv \theta(\Theta_0,0) \;,
	\label{eq:thetaConj}
\ee
in the same way that the action-angle transformation conjugates the true dynamics to rigid rotation \cite{Goldstein_80}, which is why we term $(\Theta,J)$ \emph{generalised action-angle variables}, (In field-line terms, they might also be called ``straight-pseudo-field-line coordinates''.) In the following we propose a variational method for satisfying the new QFMin Euler--Lagrange equation \eref{eq:QFMinEL} \emph{and} the reconciliation condition in the simplified form \eref{eq:SFL}, thus making the new QFMin torus coincide with the new ghost torus.

\section{Relabelling Transformation}\label{sec:relabtrnsfn}

Do the reconciliation conditions \eref{eq:QFMinEL} and \eref{eq:SFL} define the reconciliation transformation \eref{eq:thetaTransfn} uniquely? We show in this section that the answer is in the negative --- if there exists at least one solution, then there exists an \emph{infinity} of different solutions generated by a class of transformations, $\Theta \mapsto \bar\Theta$, of the form
\begin{equation}\label{eq:relabtrans}
	\bar\Theta(\Theta,\zeta) = \Theta +\tilde\Theta(\Theta - \iotabar\zeta) \;,
\end{equation}
with $\tilde\Theta(\Theta_0)$ any function of $\Theta_0$ that is $2\pi\qr$-periodic. The necessity of the $2\pi\qr$-periodicity restriction is to preserve $2\pi$-periodicity in $\zeta$, as can best be seen in Fourier representation,
\begin{equation}\label{eq:relabtransFourier}
	\bar\Theta(\Theta,\zeta) = \Theta + \sum_{m_{\rm r}}\tilde\Theta_{m_{\rm r}}\sin {m_{\rm r}}(\Theta - \iotabar\zeta) \;,
\end{equation}
(assuming odd parity, and hence a sine series) where the \emph{resonant} poloidal Fourier indices $m_{\rm r}$ are integer multiples of $\qr$ in order that corresponding resonant toroidal Fourier indices $n_{\rm r} = m_{\rm r}\iotabar \equiv m_{\rm r}\pr/\qr$ exist.

If $\Theta = \Theta_0 + \iotabar\zeta$ then $\bar\Theta  =\bar \Theta_0 + \iotabar\zeta$, where $\bar\Theta_0 \equiv \bar\Theta(\Theta_0,0)$. Thus these transformations rearrange the set of pseudo-orbits $\Theta_0 + \iotabar\zeta$ by translating them up and down in the $\zeta,\Theta$ Cartesian plane in such a way that they are all still rectilinear with slope $\iotabar$. As each $\Theta_0$ labels a different orbit after the transformation than it did before, we call transformations of the form \eref{eq:relabtrans} \emph{relabelling transformations}.

Composing relabelling transformations with the reconciliation transformation \eref{eq:thetaTransfn} gives what we now proceed to show to be an equivalence class of reconciliation transformations $\bar\theta(\Theta,\zeta)$,
\be \bar\theta(\bar\Theta(\Theta,\zeta),\zeta) \equiv \theta(\Theta,\zeta) \;.
	\label{eq:thetaTransfnrelab}
\ee
The relabelled conjugacy equation, analogous  to \eref{eq:thetaConj}, is
\be
	\bar\vartheta(\zeta|\bar\theta_0) = \bar\theta(\bar\Theta_0 + \iotabar\zeta,\zeta) \;, \:
	\bar\theta_0 \equiv \bar\theta(\bar\Theta_0,0) \;.
	\label{eq:thetaConjrelab}
\ee
the velocity $\vartheta'$ and acceleration $\vartheta''$ transform similarly. Thus, the ``unreconciled'' action gradient \eref{eq:actiongrad}  is invariant under relabelling,
\begin{equation}\label{eq:actgradinvariance}
	\frac{\delta S}{\delta\theta}(\bar\vartheta,\bar\vartheta',\zeta)
	=  \frac{\delta S}{\delta\theta}(\vartheta,\vartheta',\zeta) \;.
\end{equation}

Differentiating both sides of \eref{eq:thetaTransfnrelab} with respect to $\Theta$ gives
\be \bar\Theta_{\Theta}(\Theta,\zeta)\bar\theta_{\bar\Theta}(\bar\Theta,\zeta) \equiv \theta(\Theta,\zeta) \;,
	\label{eq:thetaTransfnrelabdiff}
\ee
where, from \eref{eq:thetaTransfnrelab}, $\bar\Theta_{\Theta}(\Theta,\zeta) = 1 + \tilde\Theta'(\Theta - \iotabar\zeta)$. Using \eref{eq:actiongrad} and \eref{eq:thetaTransfnrelabdiff} we find the relabelling transformation condition for the ``reconciled'' action gradient \eref{eq:newActionGrad},
\begin{equation}\label{eq:newActionGradbar}
	\bar\Theta_{\Theta}(\Theta_0)\frac{\delta S}{\delta\bar\Theta}
	 = \frac{\delta S}{\delta\Theta} \;.
\end{equation}
Taking the total derivative of both sides of \eref{eq:newActionGradbar} with respect to $\zeta$ and observing that $\d\bar\Theta_{\Theta}(\Theta_0)/\d\zeta = 0$ commutes with $\d/\d\zeta$, if \eref{eq:QFMinEL} is satisfied before the relabelling transformation, it will be satisfied after, and thus \emph{the relabelling transformations \eref{eq:relabtrans} generate equivalent reconciled solutions}. 

Finally, consider the case of infinitesimal relabelling transformations,
\begin{equation}\label{eq:deltalabtrans}
	\bar\Theta(\Theta,\zeta) = \Theta +\delta\tilde\Theta(\Theta - \iotabar\zeta) \;.
\end{equation}
Expanding \eref{eq:thetaTransfnrelab} to first order we find the general \emph{relabelling variation}
\begin{equation}\label{eq:thetarevar}
	\delta\theta(\Theta,\zeta) = -\delta\tilde\Theta(\Theta - \iotabar\zeta)\,\theta_{\Theta}(\Theta,\zeta) \;.
\end{equation}

\section{Variational Formulation}\label{sec:VarQFMinGhost}
We can impose condition \eref{eq:SFL} simply by using it as an ansatz for the pseudo-orbit paths in $\Theta,\zeta$ space [which then defines the paths in $\theta,\zeta$ space via \eref{eq:thetaTransfn}: $\vartheta(\zeta) = \theta(\Theta_0 + \iotabar\zeta,\zeta)$], but imposing \eref{eq:QFMinEL} is more difficult as it is nonlinear. In \cite{Dewar_Hudson_Gibson_10} we used perturbation theory, but this is limited to small departures from integrability. Instead we now introduce a variational approach that can be used to find numerical solutions for arbitrarily nonlinear problems by building on standard optimisation methods.

As the (primary) objective functional we take
\begin{eqnarray}
	F[\theta] & \equiv & \frac{1}{2} \int_0^{2\pi/\qr}\! \frac{\qr\d\Theta_0}{2\pi} \!\!\int_0^{2\pi \qr}\! \frac{\d \zeta}{2\pi\qr}
	\left[\frac{1}{\theta_{\Theta}}\left(\frac{\d}{\d\zeta} \frac{\delta S}{\delta\Theta}\right)^2\right]_{\vartheta = \theta(\Theta,\zeta),\:\Theta=\Theta_0 + \iotabar\zeta}
	\\ \nonumber
	& = &  \frac{1}{2}\!\!\!\int\limits_{\:0}^{\quad\: 2\pi}\!\!\!\!\!\!\!\int \frac{d\Theta \d\zeta}{(2\pi)^2}
	\frac{1}{\theta_{\Theta}}\left[ \frac{\d}{\d\zeta}\left(\theta_{\Theta}\frac{\delta S}{\delta\theta}\right)\right]^2
	\;,
	\label{eq:Obj1}
\end{eqnarray}
where we have obtained the second form of $F$ by changing variables from $\zeta,\Theta_0$ to $\zeta,\Theta$, so that, in the first form, $\d/\d\zeta$ denotes $\partial_{\zeta}$ with $\Theta_0$ fixed, while in the second form $\d/\d\zeta = \partial_{\zeta} +\iotabar\partial_{\Theta}$. The details of the reshuffling of the limits of the integrals are similar to those spelt out in \cite[eq. (45)]{Dewar_Hudson_Gibson_12}. Also, in the second form we have used \eref{eq:newActionGrad} to express $\delta S/\delta\Theta$ in terms of $\theta_{\Theta}(\Theta,\zeta)$ and $\delta S/\delta\theta$, which is given explicitly in terms of the Lagrangian $L$ in \eref{eq:actiongrad}. The weight factor $1/\theta_{\Theta}$ has been inserted to make $F$ precisely invariant with respect to relabelling transformations whether or not \eref{eq:QFMinEL} is satisfied, as can be demonstrating by making a change of variable from $\Theta$ to $\bar\Theta$ and using the results of \sref{sec:relabtrnsfn}.

The objective functional $F$ is to be minimized over all functions $\theta(\Theta,\zeta)$ such that $\theta(\Theta,\zeta)$ is a $2\pi$-periodic function of $\zeta$, is monotonically increasing in $\Theta$, and $\theta(\Theta+2\pi,\zeta) = \theta(\Theta,\zeta) + 2\pi$. (Or, equivalently, such that $\theta_\Theta$ is $2\pi$-periodic in both $\Theta$ and $\zeta$, its $\Theta$-average is unity, and its $\zeta$-average is zero.) Clearly $F[\theta] \geq 0$, with equality applying iff \eref{eq:QFMinEL} is satisfied.


\section{Constraining the relabelling symmetry}\label{sec:2ndOpt}

If $F = 0$ for some $\theta(\Theta,\zeta)$, an infinity of new solutions can be generated by applying finite relabelling transformations  (\ref{eq:thetaTransfnrelab}) --- \emph{the solution to the QFMin condition (\ref{eq:QFMinEL}) is not unique}. Without a further constraint to fix the solution, no optimisation algorithm for minimising $F$ can ever converge.

One such constraint method derives from the Fourier form of the relabelling transformation, \eref{eq:relabtransFourier}, where it is seen that the relabelling symmetry gives us precisely enough freedom to constrain the resonant Fourier coefficients in the expansion of $\theta$ to any desired value, thus removing the non-uniqueness in the reconciliation transformation.

This constraint approach was implicitly adopted in the perturbation method given in \cite{Dewar_Hudson_Gibson_12}, where we set the resonant Fourier coefficients in the expansion of $\theta$ to \emph{zero} at first and second order in nonlinearity. An analogous zero-resonant-coefficients constraint method was used in a nonlinear numerical study \cite{Gibson_12}, performed using the analogue of our primary objective functional $F$ for the standard map. While this study gave convincing evidence that reconciliation transformations exist for systems with quite large islands and chaotic regions (standard map nonlinearity parameter $k \sim 1$) it was found that the method broke down for stronger nonlinearity because the transformation $\Theta \mapsto \theta$ became non-invertible for $k \gg 1$.

As there is no compelling reason to set the resonant coefficients to zero, we propose that a better approach would be to determine them by minimising a \emph{secondary objective function} $G$, not invariant under relabelling, over the equivalence class generated by the relabelling transformations \eref{eq:relabtrans}. As the failure of the zero-resonant-coefficients constraint method manifested itself by $\theta_{\Theta}$ going negative, a natural choice for secondary objective functional is the average of the weight function $1/\theta_{\Theta}$,
\begin{equation}
	G[\theta] =  \frac{1}{2}\!\!\!\int\limits_{\:0}^{\quad\: 2\pi}\!\!\!\!\!\!\!\int \frac{d\Theta \d\zeta}{(2\pi)^2}
	\frac{1}{\theta_{\Theta}(\Theta,\zeta)}
	\;,
	\label{eq:Obj2}
\end{equation}
which maintains monotonicity because $G$ diverges toward $+\infty$ if $\theta_{\Theta} \to +0$ anywhere in the integration domain. Using the Schwarz inequality for the inner product between the two functions $\sqrt\theta_{\Theta}$ and $1/\sqrt\theta_{\Theta}$ we can also show that $G$ is bounded below by unity, so $1 \leq G < \infty$.

Preliminary numerical experiments on minimising the analogue of (\ref{eq:Obj2}) for the standard map to fix the resonant $\theta_m$ indicate that breakdown of invertibility can be avoided for higher $k$, but more systematic studies need to be performed, using appropriate dual-objective numerical optimisation methods, before one can conclude that reconciliation can be performed for arbitrary nonlinearity.

Unfortunately, despite the ubiquity of Lie symmetries in physics, there appear to be no algorithms in the standard numerical optimisation texts appropriate to this problem. Multi-objective optimisation problems are well known, but these involve a trade-off between competing objectives (the Pareto problem), whereas we wish to give 100\% Pareto weight to $F$ and minimise $G$ only over the subspace of directions where $F$ does not change. [Note that this nullspace is not precisely the same as the subspace spanned by the resonant Fourier modes because of the factor $\theta_{\Theta}$ in (\ref{eq:thetarevar}).]

A similar relabelling problem arises in three-dimensional numerical MHD equilibrium calculations, where the objective function is the total plasma and magnetic energy and the relabelling symmetry arises from the arbitrariness of choosing the poloidal angle within magnetic surfaces. A numerical method \cite{Hirshman_Meier_85,Hirshman_Breslau_98} for fixing the poloidal angle has been implemented in the VMEC code \cite{Hirshman_Betancourt_91}, using a measure of the width of the Fourier spectrum as the secondary objective function. Adaptation of this method, and other optimisation methods, to the present problem will be reported elsewhere.

\section{Conclusion}\label{sec:Conclusion}
We have reviewed the motivation and formulation of a recently published \cite{Hudson_Dewar_09} unification (reconciliation) of ghost and quadratic-flux-minimizing (QFMin) surfaces by transforming to a new poloidal angle, and have identified a relabelling symmetry that makes the reconciliation transformation non-unique. We have proposed a variational approach using a primary objective function to satisfy the reconciliation conditions and a secondary objective function to fix the relabelling symmetry, giving an explicit expression for the gradient of the primary objective function and identifying the nullspace of its Hessian operator (the space spanned by infinitesimal relabelling transformations).

Numerical validation of the method at high nonlinearity remains for further work. Also, the localised variational approach presented here is based on a canonical transformation tailored to reconciling the ghost and QFMin approaches on a \emph{single} resonant surface using a Lagrangian approach (a point transformation). To find a global pseudo-magnetic coordinate system \cite{Hudson_Dewar_98,Hudson_Dewar_99}, we need to generalise this approach to find a  canonical transformation that allows a simultaneous multi-surface optimisation. We anticipate that the generality of Hamiltonian methods for constructing ghost and QFMin surfaces \cite{Dewar_Hudson_Gibson_12} is better adapted to this purpose than the simple Lagrangian approach used in the present paper. 

\setcounter{section}{1}

\section*{References}
\bibliographystyle{iopart-num}
\bibliography{RLDBibDeskPapers}

\end{document}